\def\year{2026}\relax

\documentclass[letterpaper]{article}
\usepackage{aaai25} 
\usepackage{times}  
\usepackage{helvet}  
\usepackage{courier}  
\usepackage[hyphens]{url}  
\usepackage{graphicx} 
\usepackage{natbib}  
\usepackage{caption} 
\usepackage{subcaption} 
\DeclareCaptionStyle{ruled}{labelfont=normalfont,labelsep=colon,strut=off} 
\usepackage{algorithm}
\usepackage{algorithmic}
\usepackage{svg}
\svgsetup{inkscapelatex=false}
\usepackage{svg}
\usepackage{amsmath} 
\usepackage{tabularx}
\usepackage{booktabs}
\usepackage{tcolorbox}
\usepackage{enumitem}
\newcommand{\maha}{\ensuremath{\text{MAHA}}}
\newcommand{\amaha}{\ensuremath{\text{anti-MAHA}}}
\usepackage{xcolor}

\usepackage{newfloat}
\usepackage{listings}
\usepackage{color}
\usepackage{xcolor}
\usepackage{tabularx}
\usepackage{booktabs}
\usepackage{array}
\usepackage{multirow} 
\floatstyle{ruled}
\newfloat{listing}{tb}{lst}{}
\floatname{listing}{Listing}
\title{The Structure and Dynamics of the Online MAHA--sphere}
\author{
    Sabit Ahmed,
    Subigya Nepal,
    Henry Kautz
    }
\affiliations{
    bcw3zj@virginia.edu,
    sknepal@virginia.edu,
    rmw7my@virginia.edu
}
\begin{document}
\maketitle
\begin{abstract}
The ``Make America Healthy Again" (MAHA) movement has created a complex ideological ecosystem within online communities, where advocacy for healthier lifestyles and whole-food diets coexists with vaccine skepticism, anti-science attitudes, opposition to genetically modified crops and public water fluoridation. Understanding how these interconnected beliefs interact, overlap, and evolve is critical for public health communication and intervention.
Unlike previous studies that largely examined individual health-contested discourses in isolation (e.g., anti-vaccine or anti-science narratives), we uncover the functional overlaps, network structures, engagement patterns, opinion dynamics, and linguistic differences across the full spectrum of MAHA ideologies. Using large-scale Reddit data spanning six years, we identified 12 MAHA-adjacent themes, including mainstream topics such as exercise, whole food, and screen use, as well as contentious topics such as vaccines, masks, GMOs, fluoride, and others. We developed a tree-based few-shot LLM pipeline to classify stances (pro, anti, neutral) across all themes, then computed user-level opinion scores to examine cross-theme interactions and opinion shifts over time. We find that MAHA-aligned (pro-MAHA) users exhibit strong cross-theme bundling and coherent network structure, whereas anti-MAHA users do not bundle beyond chance.
MAHA users cluster in a few mainstream subreddits (70\% in r/politics alone), but post in a wide ecosystem of MAHA-related communities. 
During the COVID-19 pandemic, 25\% and 18\% of anti-fluoride and anti-mask posters transitioned into anti-vaccination posts, and more than 10\% of MAHA posters later moved from vaccine discourse to broader anti-science narratives, suggesting that vaccine skepticism may serve as an entry point into wider anti-science engagement. 
Finally, pro- and anti-MAHA communities also exhibit distinct psycholinguistic profiles, reflecting deeper ideological and rhetorical divides.
\end{abstract}

\section{Overview}

The ``Make America Healthy Again'' (MAHA) movement, a term coined by Robert F. Kennedy Jr. during the 2024 election campaign, originated as a small community of vaccine skeptics, critics of pesticides and environmental chemicals, and advocates of organic food and alternative medicine, united under the banner of ``health freedom'' \citep{stolberg2026maha}. Fueled by social media influencers and political figures, people from across the ideological spectrum later joined, and MAHA grew into a popular health reform movement.
 Although supporters of the MAHA movement underscore the importance of a healthy diet, organic food and exercise, they criticize the overuse of medication and the expansion of vaccine schedules, the impact of screen and technology use on teens, and the effect of microplastics and chemicals from pesticides and similar products. 

Social media has become a primary alternative source of health information, creating an environment in which misinformation spreads rapidly and is difficult to correct \citep{zhao_online_2022}. Poor health literacy, misinterpretation of medical data, cognitive biases, and political polarization all make online users vulnerable to erroneous beliefs and distrust of science, institutions, and conventional medical practice \citep{Yarnell2025_MAHA_Harvard,zhao_discovering_2024,paino_medical_2024}.
Previous work on platforms such as Twitter has documented well-organized misinformation communities and conspiracy networks driving vaccine hesitancy and anti-science narratives \citep{sharma_covid-19_2022,schmitz_detecting_2023,mu_vaxxhesitancy_2023}, but these studies focus on individual health-contested topics in isolation rather than the broader belief ecosystem that movements like MAHA inhabit.
Therefore, it is critical to understand how its interconnected and sometimes contradictory beliefs interact in online spaces, how they evolve and spread to other belief systems, and what psycholinguistic patterns emerge when MAHA communities communicate online.  

To address these questions, we developed a pipeline to map the structural and functional overlap in MAHA beliefs using large-scale Reddit data, a platform that hosts billions of public posts and comments, from which we drew a six-year dataset spanning January 2020 to December 2025. 
We applied a two-stage keyword search and LLM-based stance classification to identify MAHA-aligned users, then analyzed cross-theme co-engagement, network structure, engagement behavior across subreddits, longitudinal stance transitions, and psycholinguistic patterns to address the following research questions:
\begin{itemize}
    \item \textbf{RQ1}: Do MAHA (pro-MAHA) stances across health themes co-occur as a coherent belief bundle, and does anti-MAHA show the same pattern?
    \item \textbf{RQ2}: What is the structure of the internal network of MAHA communities?
    \item \textbf{RQ3}: How are MAHA communities and their posting activity dispersed across subreddits?
    \item \textbf{RQ4}: How do MAHA users move between themes over time, and do certain themes reliably draw users into others?
    \item \textbf{RQ5}: How does MAHA language differ from anti-MAHA language?
\end{itemize}
\section{Data}
We summarize the steps performed to create the underlying dataset; full details available in \citep{ahmed2026online_movement}. We collected all submissions and comments from January 2020 to December 2025, a window that begins with the COVID-19 pandemic and captures the key period of MAHA-adjacent discourse. We use \textit{post} to refer to both submissions and comments throughout the paper.
We consider 12 MAHA-related themes that span lifestyle (e.g., exercise, food),
controversial health topics (e.g., vaccines, fluoride), and broader ideological
domains (e.g., anti-science, climate skepticism).
\begin{itemize}
\item \textbf{Exercise:} Promotion of exercise and physical fitness.
\item \textbf{Food:} Processed food criticism and promotion of whole foods.
\item \textbf{GMO:} Skepticism towards genetically modified organisms and crops.
\item \textbf{Fluoride:} Opposition to public water fluoridation.
\item \textbf{Vaccine:} Criticism of vaccines (e.g., vaccine injury).
\item \textbf{Mask:} Opposition to mask mandates in the context of COVID-19.
\item \textbf{Pharma:} Criticism of industry and evidence-based medical practice.
\item \textbf{Science:} Distrust of mainstream scientific consensus and institutions.
\item \textbf{EMF:} Unproven effects of electromagnetic field (EMF) radiation.
\item \textbf{Environment:} Concerns about pollutants (e.g., microplastics, toxic chemicals, pesticides).
\item \textbf{Climate:} Skepticism of mainstream climate science.
\item \textbf{Screen:} Concerns about the health effects of phone and computer use.
\end{itemize}
We performed a 2-stage keyword search, which resulted in 234M submissions and 1.3B comments in total. 
\subsection{Classifying Theme Relevance and Stance}
Next, we applied a logistic regression classifier on sentence-BERT embedding features to filter MAHA-relevant posts from generic discussions, i.e., ``booster dose" versus ``booster seats". 
After obtaining 12 sets of thematic data, the next challenge is to identify posters' stances on each theme: \textit{pro, anti, neutral, or none}. This gives a predictive model the flexibility to assign relevant stances to multiple themes that exist for a discussion. The \textit{none} label means that the submission or comment is not related to any MAHA theme - a false positive slipped past the logistic regression filter.
For example, ``Vaccines cause autism'' indicates a \maha{} stance, whereas ``It's a lie. Vaccines don't cause autism.'' indicates an \amaha{} stance; questions such as ``Where is the evidence?'' are neutral, and unrelated text such as ``Gun violence has increased in recent years'' is labeled \textit{none}.
The prompt additionally specifies decision rules for ambiguous cases. Questions without a clear stance are labeled \textit{neutral}; questions that imply a stance are labeled \maha{} or \amaha{} accordingly. Users discussing or criticizing others' views (e.g., a skeptic describing mainstream claims) are labeled by their underlying stance, not the stance they are describing.
For comments, the parent post defines the root discussion, and the child's stance is determined relative to it.
We employed Qwen2.5-32B-Instruct (AWQ-quantized) model to infer stances.

\subsection{Scoring and Classifying Users}
After obtaining classified submissions and comments, we have determined the user-level opinion score using pro, anti, and neutral stances. Opinion scores are widely used to examine polarization in online communities \cite{10.1063/1.4913758_opinion2015,PhysRevLett.124.048301_opinion2020,Wu2025_TwitterVaccineJapan}.
For each user $u$ and stance-bearing theme $t$, we aggregate LLM-assigned
stances across all posts (submissions and comments) authored by $u$ in theme
$t$. Each post is labeled \textit{pro}, \textit{anti}, or \textit{neutral}.
Let $n_{\text{pro}}^{s}$, $n_{\text{anti}}^{s}$, $n_{\text{neutral}}^{s}$ denote stance counts from submissions, and $n_{\text{pro}}^{c}$, $n_{\text{anti}}^{c}$,
$n_{\text{neutral}}^{c}$ from comments.
However, previous studies \cite{Wu2025_TwitterVaccineJapan} leveraged Twitter posts and considered unweighted scoring. In our case, users typically author far more comments than submissions; thus, the unweighted score can be dominated by comment-level stances and may under-represent a user's explicitly stated opinions in original posts.
To ensure that submissions and comments carry equal influence regardless of how many each user authored, we normalize their contributions using
weights $w_{s}$ and $w_{c}$, where $n_{s}$ and $n_{c}$ are the total
submission and comment counts for that user and theme:  Let $n_{s} = n_{\text{pro}}^{s} +
n_{\text{anti}}^{s} + n_{\text{neutral}}^{s}$ and $n_{c} = n_{\text{pro}}^{c}
+ n_{\text{anti}}^{c} + n_{\text{neutral}}^{c}$ be the total submissions and comments, respectively. The weights are:
\begin{equation}
  w_{s} = \frac{n_{c}}{n_{s} + n_{c}}, \qquad
  w_{c} = \frac{n_{s}}{n_{s} + n_{c}}
  \label{eq:op_weights}
\end{equation}

A user with few submissions and many comments receives a high weight for their submissions, so that both post types contribute equally to the final score:
\begin{equation}
  S(u,t) =
    \frac{w_{s}\,(n_{\text{pro}}^{s} - n_{\text{anti}}^{s})
        + w_{c}\,(n_{\text{pro}}^{c} - n_{\text{anti}}^{c})}
         {w_{s}\,n_{s} + w_{c}\,n_{c}}
  \label{eq:op_score}
\end{equation}

Both scores range from $-1$ (anti-MAHA) to $+1$ (MAHA). We label users as \textit{MAHA} if $S \geq 0.2$ or \textit{anti-MAHA} if $S \leq -0.2$; users with $|S| < 0.2$ are excluded from subsequent analyses, as their net stance signals are too weak to attribute a clear position. 
\section{Analytics}
To perform analysis on user overlap and opinion dynamics, we filtered users who were active across all 6 years ($\geq1$ post per year in each of 2020--2025). This gave us a total of 45{,}599 users.
In this study, we specifically focused on MAHA and anti-MAHA users and discarded the neutral user set for the downstream analyses.
    
\subsection{User Overlap Across Themes}
\begin{figure*}[t]
\centering
\begin{minipage}[t]{0.49\textwidth}
  \centering
  \includegraphics[width=\textwidth]{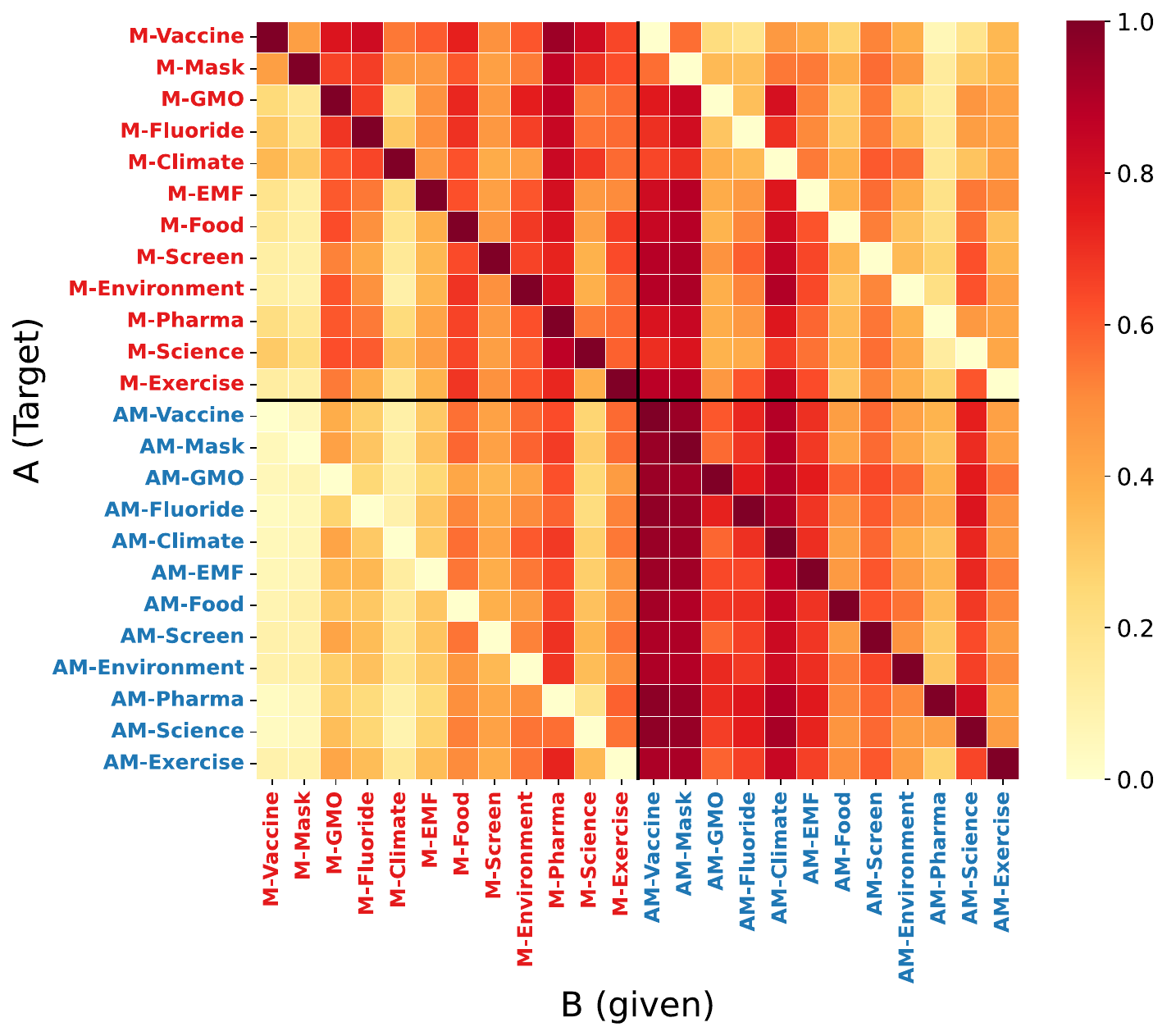}\\
\end{minipage}\hfill
\begin{minipage}[t]{0.49\textwidth}
  \centering
  \includegraphics[width=\textwidth]{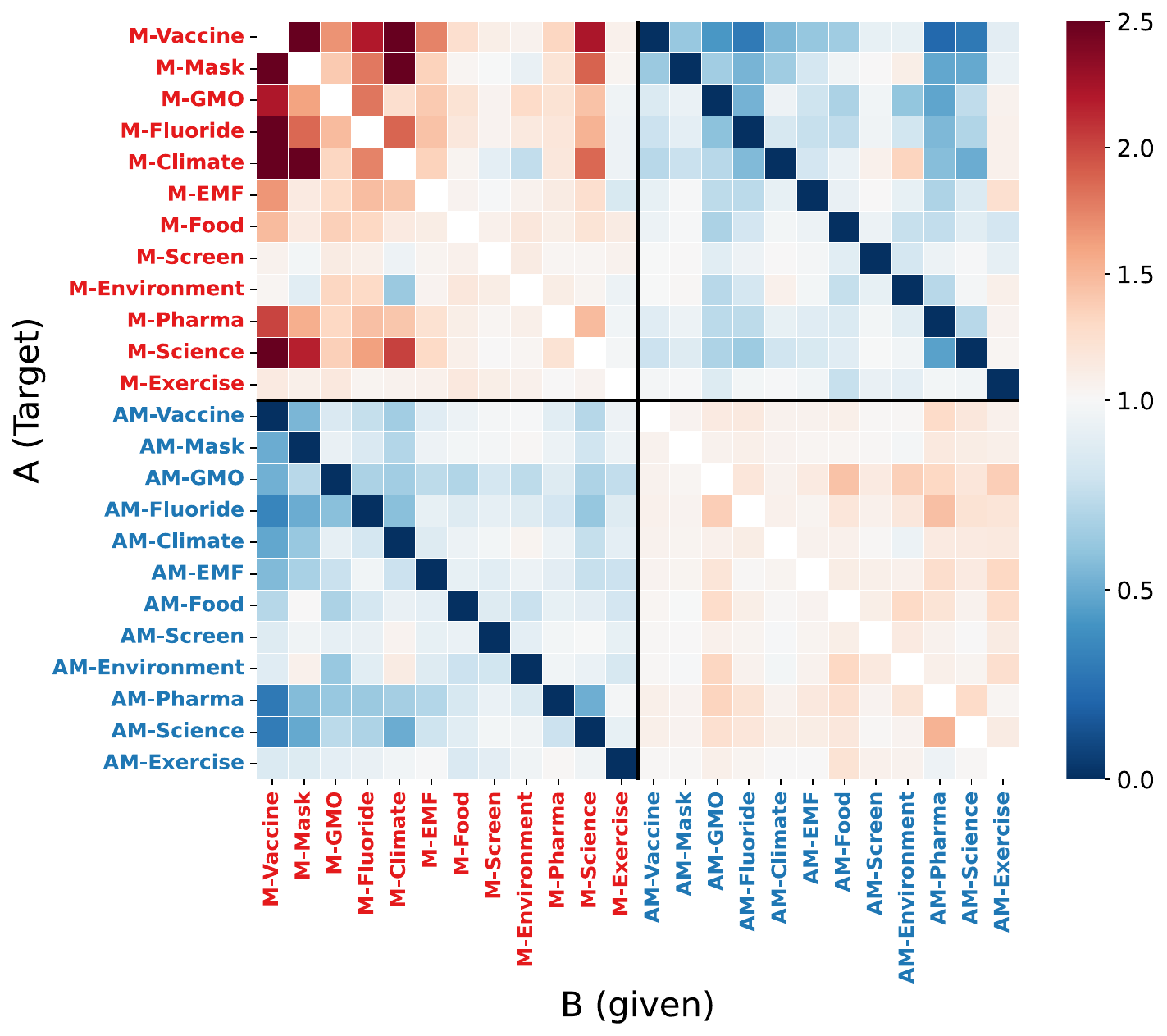}\\
\end{minipage}
\caption{
Cross-theme co-occurrence among MAHA (M) and anti-MAHA (AM) users. Both heatmaps show the MAHA theme-stance pairs (denoted by A and B); axis labels are colored red for M- pairs and blue for AM- pairs, producing four quadrants.
(A) Conditional probability --- P(A $|$ B): fraction of users in B who also appear in A. Values closer to 1 indicate stronger co-occurrence.
(B) Lift --- P(A $|$ B) / P(A): conditional probability normalized by baseline prevalence of A. Lift $> 1$ indicates higher-than-expected co-occurrence.
}
\label{fig:prob_lift}
\end{figure*}

We first examined co-engagement patterns across all theme--stance pairs.
To quantify the overlap between theme--stance pairs $A$ and $B$, we compute the conditional probability and lift as follows:
\begin{align}
P(A \mid B)
&=
\frac{|U_A \cap U_B|}{|U_B|}
\label{eq:condprob}
\\[4pt]
\mathrm{Lift}(B \rightarrow A)
&=
\frac{P(A \mid B)}{P(A)}
\label{eq:lift}
\end{align}
We treat each node $A$ (and $B$) as a theme--stance pair, with $S_A, S_B \in \{\mathrm{M}, \mathrm{AM}\}$ denoting stance labels.
The conditional probability $P(A \mid B)$ represents the fraction of users in $B$ who also appear in $A$. 
Lift quantifies how much more (or less) likely $A$ occurs among users in $B$ relative to its overall prevalence. 
Thus, a lift $> 1$ indicates that $A$ is more likely among users in $B$, lift $< 1$ indicates that $A$ is less likely, and lift $= 1$ indicates no difference from baseline, i.e., $P(A \mid B) = P(A)$ \footnote{P(A$|$B) and lift are shown jointly as each has complementary limitations (marginal prevalence vs. group-size sensitivity); scale-invariant log-OR is used for transitions analysis (Table \ref{tab:transitions}).}.
Both the conditional probability and the lift matrices are depicted in Fig \ref{fig:prob_lift} (A) and (B).
Both figures share the same layout --- the axis splits M (MAHA) from AM (anti-MAHA) stances, producing four quadrants: MAHA $\times$ MAHA (top-left), MAHA $\times$ anti-MAHA (top-right), anti-MAHA $\times$ MAHA (bottom-left), and anti-MAHA $\times$ anti-MAHA (bottom-right).
Both the top-left quadrant (within the MAHA group) and the bottom-right quadrant (within the anti-MAHA group) of Fig \ref{fig:prob_lift} (A) are dark and densely saturated. 
This pattern suggests a clear ideological divide between the MAHA and anti-MAHA groups across all themes.
After adjusting for the marginal prevalence of each theme, the lifts from Fig \ref{fig:prob_lift} (B) revealed a stark asymmetry. The M-Vaccine, M-Mask, M-Fluoride, M-Climate, and M-Science show a high lift $\geq$ 2, which means that holding a MAHA stance on these themes makes one 2-2.5$\times$ more likely than baseline to hold a MAHA stance on the others.
Lower lifts in mainstream themes like M-Food, M-Screen, M-Environment, and M-Exercise show that they barely predict each other beyond chance. 
Interestingly, the negative associations between M-Climate and M-Environment, and between M-Mask and M-Environment, suggest that MAHA users discussing environmental concerns are not necessarily climate denialists and against mask mandates.

Although anti-MAHA users show higher conditional probabilities of co-engagement, lift values of $\simeq$ 1 - 1.2 for the majority of the themes suggest that anti-MAHA users are widespread and do not hold a concrete ideological belief like MAHA users. However, this pattern should be interpreted with caution. The anti-MAHA group likely encompasses a heterogeneous population of mainstream science supporters rather than a coherent opposing movement, and their low cross-theme coherence may reflect this diversity rather than a meaningful ideological difference from MAHA users.
The cross-stance quadrants (bottom-right and top-left) are mostly blue (lift $\leq$ 1), showing that these interactions are rarer than what the base rates would predict. In other words, MAHA and anti-MAHA users do not have much overlap in either direction, showing the M-AM partition. There is no noticeable difference between the anti-MAHA mainstream and fringe themes.
Furthermore, holding a MAHA stance on Climate predicts holding an anti-MAHA stance on Environment, indicating that within MAHA, climate skepticism and environmental concerns do not consistently co-occur.

Although effect sizes are small among the MAHA themes, ideological overlaps are strong across fringe themes (i.e., Mask, Science, Fluoride, Climate). Even with stricter content moderation imposed by the Reddit platform, MAHA users hold a strong overlap in fringe beliefs, with some contradictions with mainstream themes. A small fraction of MAHA users potentially engage and dominate discussions in multiple themes with distinct ideologies. In contrast, with large effect sizes, anti-MAHA user overlap across themes is almost uniform. 
This suggests that despite having an anti-MAHA sentiment, users in this axis are loosely coupled and widespread across different themes.
\subsection{MAHA Theme Network}
To further illuminate the structural overlap among the MAHA communities, we created a theme-theme network based on shared user participation. We applied Kamada--Kawai, a distance-based force-directed layout, which positions nodes such that strongly connected themes (i.e., with high overlap) are placed closer together, while weakly connected themes are pushed further apart, as depicted in Fig \ref{fig:fig_network}. 
Each node represents a thematic category, and node size is proportional to the number of users expressing a MAHA stance within that theme.
Edges between themes are defined using the overlap of users who are active in both themes. For each pair of themes $A$ and $B$, we first compute a normalized overlap score:

\begin{figure}[t]
\includegraphics[width=\columnwidth]{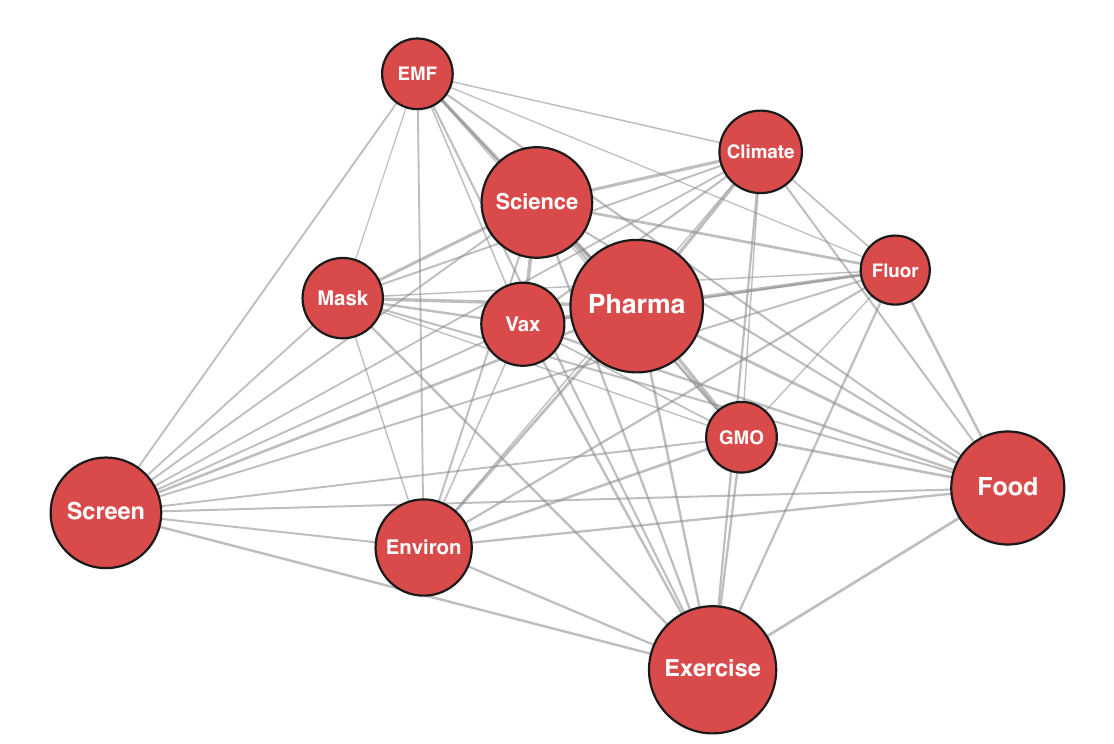}
\caption{
Distance-based layout (Kamada--Kawai) of the theme--theme interaction network among MAHA users. Node size is proportional to the number of MAHA users in each theme; edge weights are user overlaps converted to distance via Eq. 5. Themes with high overlap sit closer together.
} \label{fig:fig_network}
\end{figure}
\begin{equation}
\begin{gathered}
w_{A,B} = \frac{|U_A \cap U_B|}{\min(|U_A|, |U_B|)}, \quad
v_{A,B} = \frac{w_{A,B}}{\min_{i,j} w_{i,j}}, \\
d_{A,B} = \frac{1}{\left(v_{A,B}\right)^2}
\end{gathered}
\label{eq:kk_d}
\end{equation}
where $U_A$ and $U_B$ denote the sets of users expressing a MAHA stance on themes $A$ and $B$, respectively. 
The quantity $w_{A,B}$ measures the fraction of users that overlap relative to the smaller community. 
We then rescale weights by the smallest non-zero overlap and convert them into distances using an inverse-square transformation for layout generation.
From Fig \ref{fig:fig_network}, we can observe that Pharma, Vaccine and Science are the 3 closest nodes that potentially signal a hub-like pattern, connecting 5 other skeptical themes like Mask, EMF, Climate, Fluoride, and GMO.
While contentious themes are closely connected with each other at the center, mainstream communities vocal on Screen, Exercise, and Food are positioned at the periphery of the network. 
\subsection{Hierarchical Clustering}
\begin{figure} [ht]
\includegraphics[width=\columnwidth]{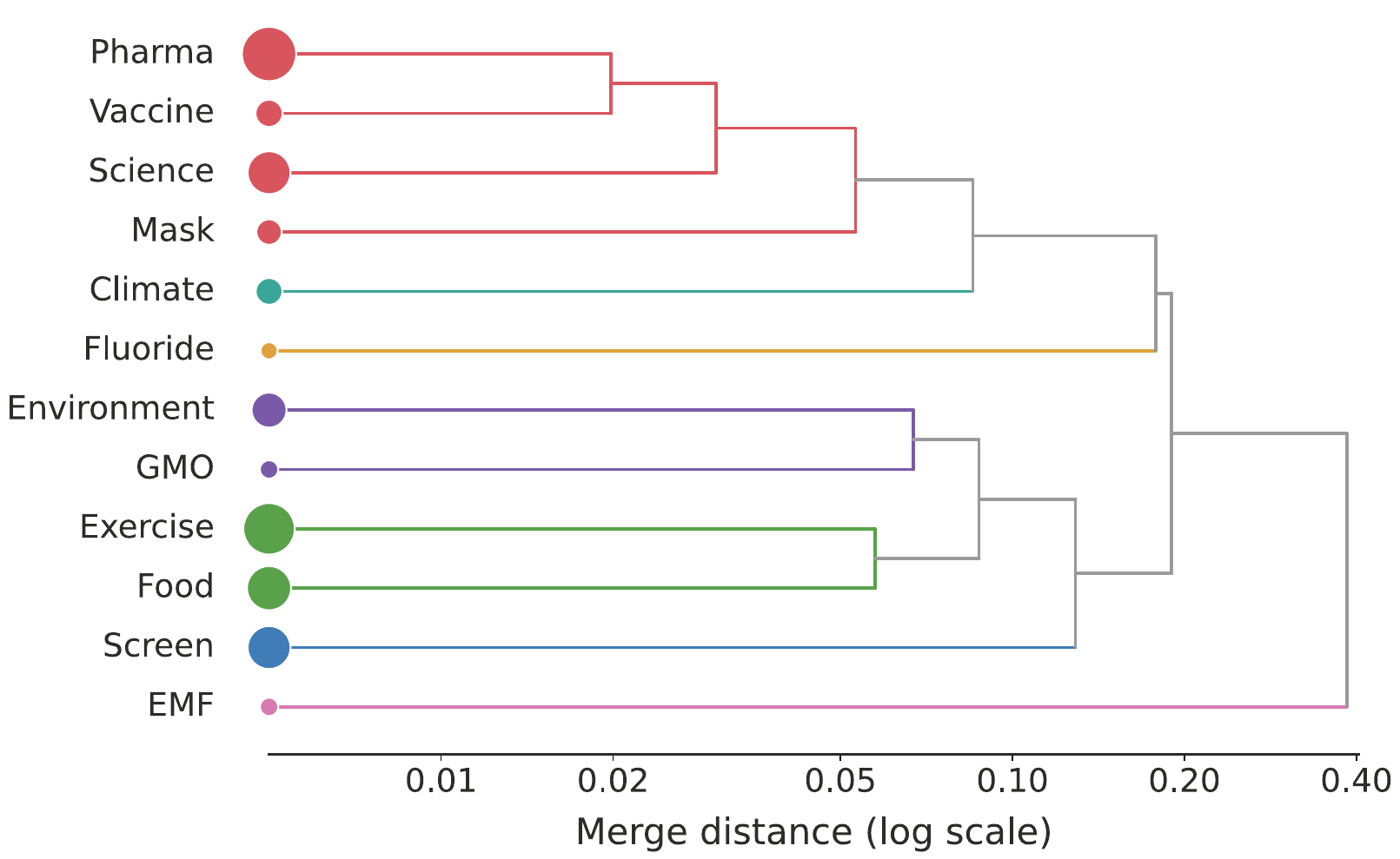}
\caption{
Hierarchical clustering (UPGMA) of MAHA themes.
The log-scale x-axis shows the merge distance (derived from Eq. \ref{eq:kk_d}) at which each pair or cluster joins; smaller values indicate stronger user overlap. Node size is proportional to the number of MAHA users in each theme.
} \label{fig:fig_cluster}
\end{figure}

We also clustered 12 themes based on how much the MAHA user base overlaps. We applied a bottom-up hierarchical clustering method named Unweighted Pair Group Method with Arithmetic Mean (UPGMA) using the derived distances from Eq. \ref{eq:kk_d} between each pair of themes.
Node area is proportional to the MAHA user counts for each theme. The x-axis shows the merge distance of nodes on a log scale.
At each step, UPGMA merges the two closest clusters; the merge distance of a join is the distance between the merged clusters at that step.
Fig \ref{fig:fig_cluster} shows 3 primary clusters merged at $\sim$0.07 (colored as red, green, and violet respectively). The red cluster contains the Pharma, Vaccine, Science, and Mask discourses, which exhibit institutional distrust and medical skepticism. The green cluster contains 2 mainstream themes: Exercise and Food. The violet cluster with Environment and GMO merged right after the formation of the mainstream and institution distrust clusters, showing that these themes might act as a bridge between the distrust and lifestyle-oriented/mainstream MAHA groups. Climate, Screen, and Fluoride joined at a later merge step $~0.1 - 0.2$, while EMF was the last element merged at $\sim$0.35.

\begin{figure*}[t]
\centering

\begin{subfigure}[t]{0.49\textwidth}
    \centering
    \includegraphics[width=\linewidth]{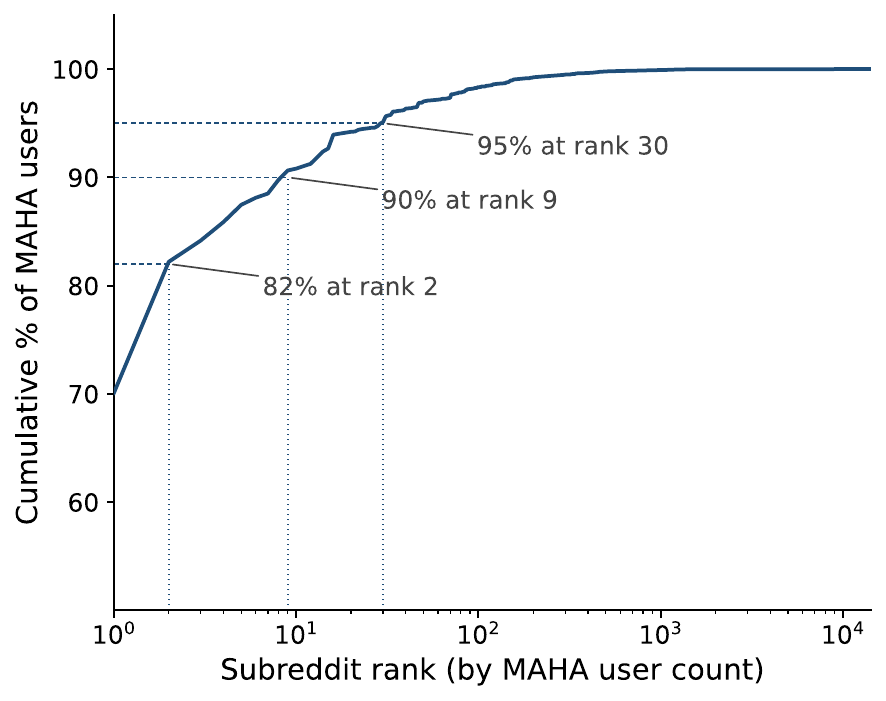}
\end{subfigure}
\hfill
\begin{subfigure}[t]{0.49\textwidth}
    \centering
    \includegraphics[width=\linewidth]{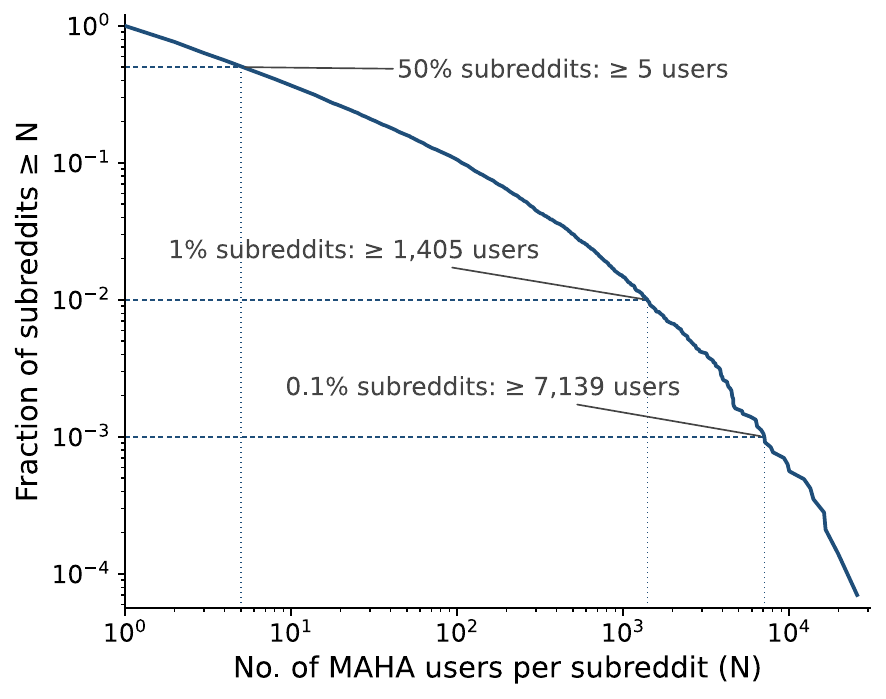}
\end{subfigure}

\caption{(A) Cumulative share of MAHA users captured by top-ranked subreddits, where subreddits are ranked by MAHA user count. (B) Complementary cumulative distribution function (CCDF) of MAHA user counts per subreddit, illustrating a heavy-tailed participation distribution across subreddits.}
\label{fig:ccdf_combined}

\end{figure*}
\subsection{Subreddit Analysis}
To assess whether \maha{} users cluster in a small number of communities or are scattered across Reddit, we computed a cumulative coverage curve as shown in Fig. \ref{fig:ccdf_combined} (A). All 14,256 subreddits containing at least one \maha{} user were ranked in descending order by their \maha{} user count. For each rank N, we computed the cumulative percentage of unique \maha{} users present in at least one of the top-N subreddits.
The curve rises steeply and flattens quickly. About 70\% of the 36,803 unique MAHA users are captured by r/politics alone; 82\% are captured by the top 2 subreddits (r/politics and r/AskReddit); and 95\% are captured by the top 30 subreddits. The remaining $\sim$14,200 subreddits collectively contribute fewer than 5\% of unique users.
Fig.~\ref{fig:ccdf_combined} (B) reports the complementary cumulative distribution (CCDF) of \maha{} users per subreddit. The near-linear decay on log-log axes indicates a heavy-tailed distribution: a small number of subreddits host thousands of \maha{} users each, while the majority host only a few.
\begin{figure} [ht]
\includegraphics[width=\columnwidth]{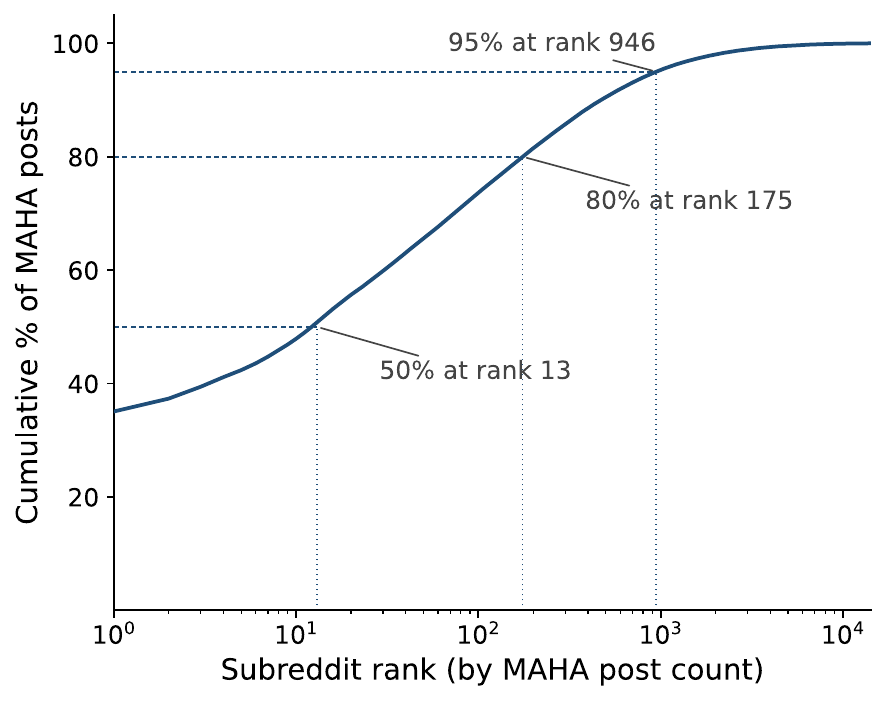}
\caption{Cumulative share of MAHA posts captured by top-ranked subreddits, where subreddits are ranked by MAHA post count.} \label{fig:fig_ccdf_post}
\end{figure}

\begin{table*}[t]
\centering
\small
\setlength{\tabcolsep}{6pt}
\renewcommand{\arraystretch}{1.25}

\begin{tabularx}{\textwidth}{>{\bfseries}p{3cm} X}
\toprule
Theme & Top-3 subreddits (fraction of MAHA posts) \\
\midrule

Exercise &
PetiteFitness30plus (0.59\%), xxfitness (0.37\%), WellnessOver30 (0.37\%) \\

Food &
ScientificNutrition (0.66\%), Microbiome (0.57\%), SaturatedFat (0.35\%) \\

GMO &
GMOMyths (0.48\%), organic (0.12\%), actualconspiracies (0.02\%) \\

Fluoride &
westmidlands (0.05\%), SantaMaria (0.04\%), water (0.03\%) \\

Vaccine &
antivaccine2 (3.01\%), unvaccinated (1.57\%), DebateVaccines (1.01\%) \\

Mask &
NorCalLockdownSkeptic (0.35\%), ParentingWithoutFear (0.35\%), CoronavirusNY (0.19\%) \\

Pharma &
antivaccine2 (0.93\%), DebateVaccines (0.34\%), LockdownCriticalLeft (0.28\%) \\

Science &
DebateVaccines (0.19\%), redditsecurity (0.19\%), futurologyappeals (0.19\%) \\

EMF &
invisiblerainbow (0.05\%), HankAaronAward (0.02\%), 5GDebate (0.01\%) \\

Environment &
bioethics (0.22\%), GMOMyths (0.14\%), organic (0.10\%) \\

Climate &
climateskeptics (0.67\%), futurologyappeals (0.47\%), climatechange (0.23\%) \\

Screen &
ReplikaTech (0.17\%), TangleNews (0.11\%), coursivofficial (0.08\%) \\

\bottomrule
\end{tabularx}

\caption{Top-3 subreddits per theme ranked by the fraction of MAHA posts among all posts in the subreddit (2020--2025).}
\label{tab:top_subreddits}

\end{table*}

While users themselves are clustered on a few subreddits, their posting \emph{activity} is distributed throughout the platform (Fig.~\ref{fig:fig_ccdf_post}).
MAHA posts are far more dispersed in aggregate: 50\% of MAHA posts are captured only by the top 56 subreddits, 80\% by the top 314, and 95\% by the top 1,283 -- roughly 50- to 150-fold more communities than are needed to capture the same fraction of users.
At the user level, the top of the ranking is dominated by large mainstream subreddits; however, they do not dominate the post distribution to the same degree: \maha{} users participate in r/politics and r/AskReddit (82\% of user coverage; 50.8\% of post coverage), but they also post extensively in MAHA-specific communities, e.g., r/conspiracy (rank 2), r/DebateVaccines (rank 3), r/LockdownSkepticism (rank 9), r/climateskeptics (rank 24), etc.

Beyond concentration patterns, Table \ref{tab:top_subreddits} reports the top three subreddits ranked by the fraction of MAHA posts (from all 45{,}599 active users) out of all posts made between 2020 and 2025 (i.e., before any keyword or relevance filtering). Vaccine discourse has the highest MAHA-share across all themes, with three vaccine-skeptic subreddits -- r/antivaccine2 (3.01\%), r/unvaccinated (1.57\%), and r/DebateVaccines (1.01\%). Other contested themes have their own niche communities: GMO skepticism in r/GMOMyths (0.48\%), Mask opposition in r/NorCalLockdownSkeptic and r/ParentingWithoutFear, Climate denial in r/climateskeptics (0.67\%), and EMF in r/invisiblerainbow and r/5GDebate -- forming an ecosystem of theme-specific MAHA communities. 
\subsection{Opinion Dynamics}
After examining the topology between MAHA and anti-MAHA users, we further analyzed the opinion dynamics of the MAHA users. 
We measured cross-theme stance transitions among MAHA users active across consecutive years. For each user, we re-computed yearly opinion scores from annual post counts using Eq.~\ref{eq:op_score}. To track how users move between themes year over year, we needed to assign each user a primary theme per year as their main focus. Because users may engage with multiple themes in a given year, we defined a user's primary theme in year $Y{+}1$ as the theme on which they posted most frequently (submissions and comments combined), with ties broken by opinion score magnitude. We then computed, for each source theme $A$ in year $Y$, the fraction of MAHA users whose primary engagement in year $Y{+}1$ was theme $B \neq A$. To identify statistically significant transitions beyond chance, we applied Fisher's exact test for each source--destination theme pair independently, constructing a $2{\times}2$ contingency table of users from theme $A$ versus all other source themes, crossed with primary destination being theme $B$ versus any other theme. The resulting $p$-values were jointly corrected for multiple comparisons within each year-pair using the Benjamini-Hochberg FDR procedure. We retained transitions with BH-FDR $<$ 0.05, $\log(\text{OR}) > 0.5$, transition rate $\geq$ 5\%, and at least 10 observed users, which are reported in Table \ref{tab:transitions}. 
\begin{table*}[ht]
\centering
\begin{tabular}{|c|l|l|c|c|c|}
\hline
\textbf{Year} & \textbf{Source} & \textbf{Destination} & \textbf{N (obs)} & \textbf{\% of Source} & $\boldsymbol{\log(\mathrm{OR})}$ \\
\hline

\multirow{3}{*}{2020$\to$2021}
& Fluoride & Vaccine & 28  & 25.2 & $1.192^{***}$ \\
& Mask     & Vaccine & 344 & 18.3 & $0.849^{***}$ \\
& Climate  & Vaccine & 174 & 14.4 & $0.512^{***}$ \\
\hline

\multirow{3}{*}{2021$\to$2022}
& Mask     & Vaccine & 342 & 15.6 & $0.588^{***}$ \\
& Fluoride & Vaccine & 26  & 16.0 & $0.571^{*}$ \\
& Vaccine  & Science & 198 & 6.8  & $0.520^{***}$ \\
\hline

\multirow{4}{*}{2022$\to$2023}
& Fluoride & Vaccine & 25  & 12.0 & $0.623^{*}$ \\
& Mask     & Vaccine & 242 & 11.1 & $0.581^{***}$ \\
& Climate  & Vaccine & 166 & 11.1 & $0.565^{***}$ \\
& Climate  & Science & 125 & 8.4  & $0.527^{***}$ \\
\hline

\multirow{6}{*}{2023$\to$2024}
& Fluoride & Vaccine & 27  & 14.2 & $1.389^{***}$ \\
& Mask     & Vaccine & 119 & 9.2  & $0.938^{***}$ \\
& Climate  & Science & 200 & 11.6 & $0.659^{***}$ \\
& Vaccine  & Science & 292 & 11.1 & $0.630^{***}$ \\
& Mask     & Science & 136 & 10.5 & $0.537^{***}$ \\
& GMO      & Food    & 36  & 9.6  & $0.515^{*}$ \\
\hline

\multirow{5}{*}{2024$\to$2025}
& Mask      & Vaccine & 86  & 7.2  & $0.789^{***}$ \\
& Fluoride  & Vaccine & 44  & 7.2  & $0.777^{**}$ \\
& Climate   & Science & 192 & 11.4 & $0.569^{***}$ \\
& Climate   & Vaccine & 98  & 5.8  & $0.561^{**}$ \\
\hline

\end{tabular}
\caption{
Significant cross-theme transitions among MAHA users (2020--2025). A transition counts a user whose primary theme switches from source to destination between consecutive years. Columns report the observed user count, the percentage of source-theme users transitioning, and the log-odds ratio. Filtered to transitions with log OR $>$ 0.5, transition rate $\geq$ 5\%, and $\geq$ 10 users.
$^{*}p<0.05$, $^{**}p<0.01$, $^{***}p<10^{-5}$ (BH-adjusted).
}
\label{tab:transitions}
\end{table*}
Table \ref{tab:transitions} shows a noticeable pattern during the COVID-19 pandemic and the post-COVID era.
Fluoride and Mask users consistently shifted to Vaccine discourse throughout the six-year timeline, with a maximum of 25\% and 18\% shift occurring during the pandemic (2020-21).
Climate was also a significant feeder to vaccine discussions for 3 distinct year pairs (2020-21, 2022-23, and 2024-25).

After COVID, the vaccine debate itself shifted to the \quad (anti-) Science block in 2021-22 and 2023-24, with 6.8\% and 11\% of users transitioning.
Climate deniers also contributed to the scientific (or pseudo-scientific) claims across three-year pairs.
More than 100 users against Mask mandates contributed to Vaccine and Science arguments simultaneously in 2023-24.
We found only one year pair (2023-24) when GMO skeptics moved to a mainstream debate, e.g., Food (position on whole food).
Table \ref{tab:transitions} highlights our previous findings that there is a dense alignment between Vaccine, Science, Mask, Fluoride, and Climate themes, where Vaccine and Science acted as sink themes. Although not quite contradictory, skeptical themes like GMO blended into MAHA beliefs through the discussion on mainstream themes, e.g., Food, as depicted in Fig. \ref{fig:fig_network} and \ref{fig:fig_cluster}. This indicates that the MAHA-sphere is an evolving space, spreading its reach beyond contentious topics to the mainstream.
\subsection{Psycholinguistic Differences Between MAHA and Anti-MAHA Discourse}
To compare the linguistic profiles of MAHA and anti-MAHA users across themes, we scored all posts of at least five words (1.9M submissions and comments). These were scored across 18 LIWC-22 \citep{boyd2022liwc22} psycholinguistic dimensions commonly used in social science and mental health research \citep{pendse2025rolepartisanculturemental,henderson2013mental}. Following \citep{pendse2025rolepartisanculturemental}, we applied per-theme 1:1 propensity score matching on four covariates: posts per year, number of active themes, mean word count, and mean function-word percentage, yielding 42{,}481 matched users across 12 themes. 
Group differences on each LIWC dimension were tested using Welch's two-sample t-test with Benjamini-Hochberg FDR correction. 
Effect sizes are reported as Cohen's d (positive = pro users score higher). Significance tiers are indicated as: $^{*} p < 0.05,\ ^{**} p < 10^{-5},\ ^{***} p < 10^{-20},\ ^{\dagger} p < 10^{-100}$. 
We interpret the matrix alongside the
illustrative example posts in Table \ref{tab:maha_example_texts}.
The most consistent pattern across the matrix is an asymmetry in Cognitive\_Processing and Feel scores. MAHA users scored significantly lower on Cognitive\_Processing (e.g., ``think'', ``know'', ``because'') than anti-MAHA users in 8 of 12 themes, with the strongest effects on GMO, Exercise, Food, Pharma, and Science. In parallel, MAHA users scored higher on Feel (e.g., ``feels'', ``touch'') --- the sensory and perceptual dimension --- on Climate, Food, Screen, Environment, Pharma, and Exercise. This suggests that MAHA engagement with contested themes such as GMO, Pharma, and Science relies less on reasoning and analytical framing and more on personal feeling and perception as justification. 
Anti-MAHA users, by contrast, consistently invoke evidence and verification. 
\begin{figure*}[t]
\centering
\includegraphics[width=0.90\textwidth]{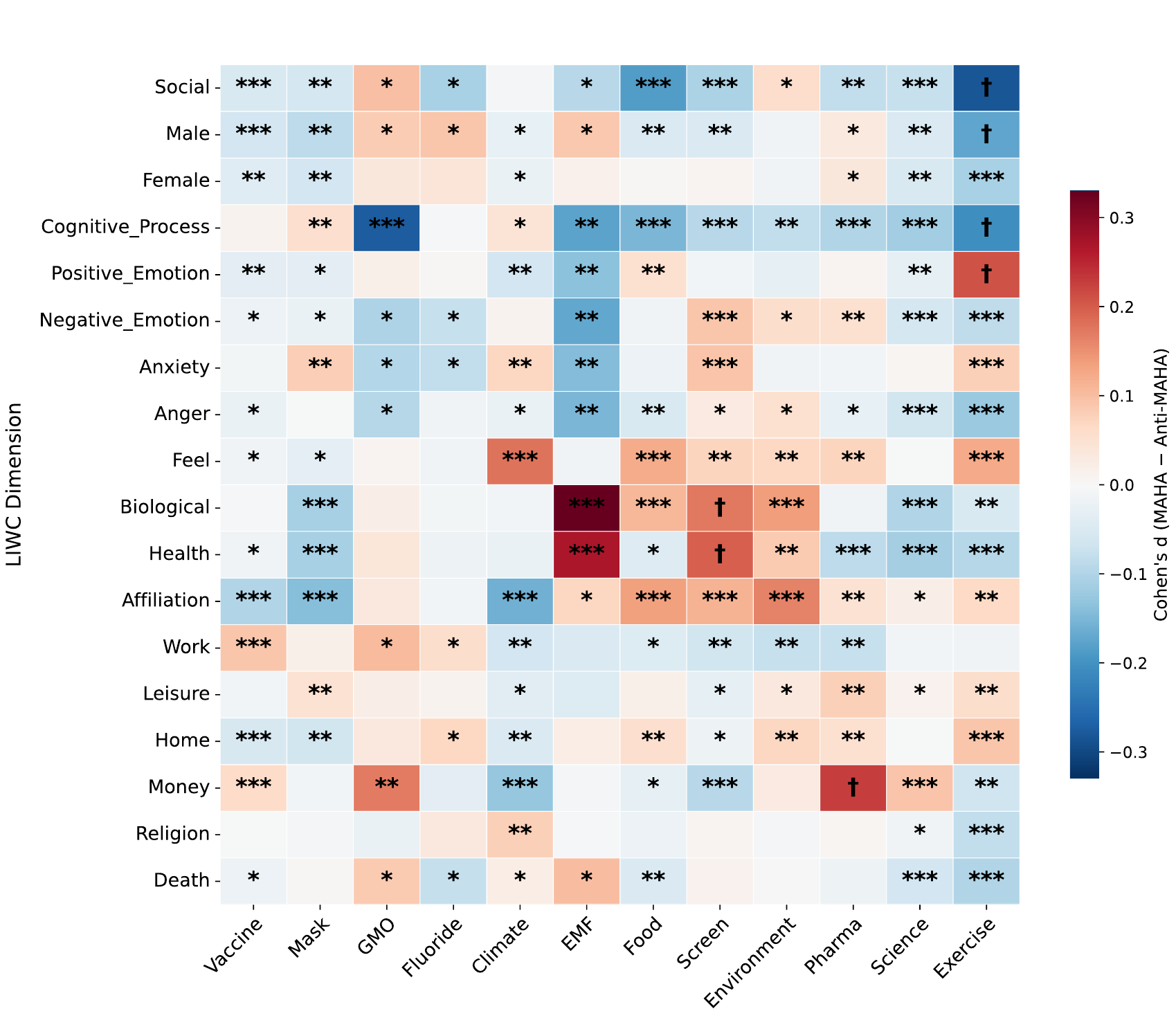}
\caption{Cohen's $d$ effect sizes for 18 selected LIWC-22 psycholinguistic dimensions comparing propensity-matched MAHA and anti-MAHA users, with themes as columns and LIWC dimensions as rows. Positive values (red) indicate higher scores among MAHA users; negative values (blue) indicate higher scores among anti-MAHA users. 
$^{*}p<0.05$, $^{**}p<10^{-5}$, $^{***}p<10^{-20}, ^{\dagger}p<10^{-100}$ (BH-adjusted).
} 
\label{fig:fig_liwc}
\end{figure*}
The Biological and Health dimensions --- which capture physical and somatic language (e.g., ``pain'', ``flu'', ``body'') --- are elevated in MAHA content for EMF, Food, Screen, and Environment. These themes concern substances, radiation, or signals that enter or act on the body, and MAHA users frame them in terms of bodily harm and physiological risk rather than abstract or policy terms. This pattern aligns with the alternative-lifestyle cluster identified in the earlier structural analysis (UPGMA), indicating that these themes share not only overlapping users but a common way of talking --- one centered on bodily integrity and what crosses it.

Money scores are higher among MAHA users on Vaccine, GMO, Pharma, and Science. 
Example posts (Table~\ref{tab:maha_example_texts}) illustrate the distrust in corporations (agriculture, food, pharmaceuticals) and established science --- e.g., \emph{``It's all a big money-making game at the expense of Americans.''} and \emph{``When reddit says `trust the science' they mean trust the science I like and have cherry picked.''} These themes form the institutional-distrust cluster in the structural analysis, and the LIWC evidence shows they also share a common framing: MAHA users talk about power and economic incentives, while anti-MAHA users talk about evidence and verification. Together with the previous pattern, this indicates that the MAHA population is not linguistically uniform --- the two structural clusters correspond to two distinct ways of writing about health.

Emotional patterns are theme-specific rather than uniform across MAHA content. MAHA users show more Positive Emotion on Exercise, higher Negative Emotion on Screen, Environment, and Pharma, more Anxiety on Mask, Climate, Screen, and Exercise, and more Anger on Screen and Environment. Social language is elevated in MAHA content on only 2 themes (GMO and Environment), suggesting limited people-oriented framing elsewhere. On Affiliation, Work, Leisure, and Home, MAHA users score higher in the lifestyle themes (Food, Environment, Exercise), consistent with a community organized around daily practice and private life rather than politics or institutions. Two additional theme-specific observations: MAHA users on EMF and GMO invoke Death more frequently than their anti-MAHA counterparts, consistent with a catastrophic or existential framing of technological and agricultural risk.
\begin{table*}[ht]
\centering
\begin{tabularx}{\textwidth}{
|l
|>{\raggedright\arraybackslash}X
|>{\raggedright\arraybackslash}X|
}
\hline
\textbf{Theme} & \textbf{MAHA example} & \textbf{Anti-MAHA example} \\
\hline

Vaccine &
\textcolor{red}{``The vaccine doesn't make immune \dots it just lowers severe illness.''} &
\textcolor{blue}{``I'm vaccinated \dots we are all in this together.''} \\
\hline

Mask &
\textcolor{red}{``There is absolutely no evidence that wearing a mask outside stops the spread.''} &
\textcolor{blue}{``If I wear a mask and then sneeze, my mask is catching all of that.''} \\
\hline

GMO &
\textcolor{red}{``The government prevents industries from making insulin because of lobbyists\dots It's all a big money-making game at the expense of Americans.''} &
\textcolor{blue}{``Monsanto will get laws enforced that benefit corporations but harm people -- how do we fight something like that?''} \\
\hline

Fluoride &
\textcolor{red}{``People have become so much easier to control after we started putting fluoride in the water.''} &
\textcolor{blue}{``You would die from water intoxication before fluoride toxicity at those levels.''} \\
\hline

Climate &
\textcolor{red}{``You can't solve climate change. It will occur and has occurred before humans were even around \dots''} &
\textcolor{blue}{``\dots we only have a very short amount of time to keep the Earth from going up 3 degrees and then we're all toast.''} \\
\hline

EMF &
\textcolor{red}{``Concerns about vaccinations, 5G wireless technology, and government overreach into people's lives.''} &
\textcolor{blue}{``A cell tower's power output is typically between 100 and 1000 watts \dots there's no way it could hurt you.''} \\
\hline

Food &
\textcolor{red}{``Nuts are minimally processed and a good option for snacking.''} &
\textcolor{blue}{``The soda tax reduced soda purchases by over 50\%.''} \\
\hline

Screen Use &
\textcolor{red}{``If social media didn't exist Trump wouldn't have won \dots we know how powerful the propaganda was.''} &
\textcolor{blue}{``You really need to detach from social media. There are other things out there -- pick up a hobby or something.''} \\
\hline

Environment &
\textcolor{red}{``Non-stick cookware is legitimately poisonous and wouldn't be legal to sell if industries didn't own the agencies tasked with regulating them.''} &
\textcolor{blue}{``It isn't too late to fix air pollution \dots Based upon humanity's historic responses, I beg to differ.''} \\
\hline

Pharma &
\textcolor{red}{``Why are they using drugs to lose weight? Diet and exercise should be enough.''} &
\textcolor{blue}{``Naturopaths are snake oil salesmen \dots you deserve real medical support.''} \\
\hline

Science &
\textcolor{red}{``When reddit says `trust the science' they mean trust the science I like and have cherry picked.''} &
\textcolor{blue}{``Tiktok is rife with misinformation. At least on reddit you can post your sources and people will downvote blatant misinformation.''} \\
\hline

Exercise &
\textcolor{red}{``Exercise, wash hands, vitamin C, D, multivitamin, plus quercetin and zinc if I get even a sniffle.''} &
\textcolor{blue}{``Exercise sucks \dots you just do it anyway because it needs done.''} \\
\hline

\end{tabularx}
\caption{Illustrative example comments for each theme. MAHA examples are shown in red and anti-MAHA examples in blue.}
\label{tab:maha_example_texts}
\end{table*}
Taken together, MAHA and anti-MAHA differ not only in what they claim or distrust, but in how they justify those claims --- through personal feeling and experience on the MAHA side versus through evidence and reasoning on the anti-MAHA side. Within MAHA, the psycholinguistic evidence independently recovers the two structural clusters identified in the earlier analysis: a bodily-harm framing in the alternative-lifestyle cluster (EMF, Food, Environment, Screen) and an institutional-distrust framing in the core contested-health cluster (Vaccine, GMO, Pharma, Science). These are distinct ways of writing within the same movement, rather than a single unified voice.
\section{Related Work}
Interrelated psychological, political, cognitive, and socio-economic processes play a critical role in online social networks, where various narratives and sentiments frequently emerge \citep{zhao_discovering_2024}. Vaccine hesitancy is identified as one of the top ten threats to global health \citep{dalege_using_2022}, and public opinions on masks during the pandemic were highly politicized \citep{Lang2021MaskOnMaskOff}. Disagreements and conflicting points of view with polarized lenses shaped people's attitudes towards science and institutions during the COVID-19 pandemic \citep{tyson_partisan_2020,lyu_social_2022,green_elusive_2020}, and partisanship has been shown to correlate with anti-science attitudes on Twitter \citep{rao_political_2021}.
Skepticism about science, medicine, and institutions has expanded to a wider spectrum of beliefs: 57\% of Americans are skeptical about the safety of genetically modified crops, 67\% believe scientists lack a clear understanding of GM crop health effects \citep{rainie_public_2015,kennedy_new_2016}, and roughly a third perceive science as a negative factor on the quality of food and environment \citep{rainie_public_2015}. Social media is also debated as a possible driver of mental health decline in American children \citep{nap2024socialmedia,odgers_great_2024,haidt_yes_2024}.

Many studies have used a combination of natural language processing (NLP) and social network analysis (SNA) to understand sentiments, and opinion shifts, misinformation spread, ideological and rhetorical divides as well as to uncover complex interconnected online network and community structures \citep{kim_capturing_2025,cetinkaya_cross-partisan_2025,green_how_2025,zhao_discovering_2024,dalege_using_2022,champion2008health,nutbeam2000health,taggart2012systematic,paino_medical_2024,tyson_partisan_2020,jiang_polarization_2021,pendse2025rolepartisanculturemental}. 
\citet{Lee2025SemanticEmbedding} proposed an LLM-based belief embedding method to investigate how differences between individual beliefs shape collective social beliefs and how an individual approaches decision-making with contradictory beliefs.
To better understand anti-vaccine sentiment, \citet{schmitz_detecting_2023} used sentence embeddings to identify users who are likely to spread anti-vaccine narratives.
\citet{Wu2025_TwitterVaccineJapan} found that Twitter communities are associated with users' opinion shifts towards the COVID-19 vaccine in Japan. 
To find the answer to how narratives evolve online and how to capture such narrative shifts, \citet{zhao_discovering_2024} used a hierarchical method that was capable of capturing collective narrative shifts in major events like COVID-19.
\citet{mitra_understanding_2016} examined Twitter users' expressions towards vaccination and identified 3 groups: consistent pro-vaccine users, consistent anti-vaccine users, and users who joined the anti-vaccine pool from similar controversial discussions. 
\citet{cetinkaya_cross-partisan_2025} investigated topics and stances associated with cross-partisan interactions on Twitter.  
Recently, \citet{alba2026maha} leveraged pre-trained language models (PLMs) to understand sentiments and topics commonly discussed in X's (former Twitter) MAHA community. Our work extends this line by examining MAHA as a multi-theme belief ecosystem on Reddit, characterizing cross-theme bundling, network topology, six-year opinion dynamics, and psycholinguistic differences across 12 themes.
\section{Conclusion and Future Work}
We presented a large-scale study of how MAHA-aligned beliefs are organized, evolve, and are expressed in online discourse. Using a tree-based few-shot LLM pipeline applied to six years of Reddit data (2020--2025), we classified user stance on 12 health-contested themes, and analyzed the \maha{} movement at five levels: user-level overlap, network topology, subreddit-level engagement, temporal dynamics, and language. 

Five findings stand out. 
First, MAHA stances (pro-MAHA) form a coherent belief cluster around public health and science-contested themes (e.g., Vaccine, Mask, Science, Fluoride, Climate). Users holding one MAHA-aligned position are more likely to hold others, while anti-MAHA users co-occur only through their shared mainstream stance, with no thematic bundling beyond what base rates predict. This structural asymmetry suggests that MAHA functions as an organized ideological movement, 
while the anti-MAHA (mainstream) side is a broader, more heterogeneous population engaged with controversial topics without comparable ideological cohesion.
Second, the network of MAHA themes shows clear internal structure: themes related to institutional distrust (Pharma, Vaccine, Science, Mask) form a central hub, while healthy-lifestyle themes (Screen, Exercise, Food) sit at the periphery. 
Third, MAHA users cluster in a few mainstream subreddits (70\% in r/politics) but post far more widely, engaging with both mainstream and MAHA-specific communities (e.g., r/conspiracy, r/DebateVaccines).
Fourth, the MAHA movement consolidated through directional cross-theme migration: Fluoride, Mask, and Climate science skepticism consistently diffused into broader Vaccine and Science skepticism over six years, and over 10\% of vaccine-skeptic users further shifted into anti-science engagement in recent years.
Finally, MAHA and anti-MAHA discourse differed in how claims are expressed and justified. MAHA users share more perceptual and experiential words and criticize power and monetary dynamics, while anti-MAHA users rely more on evidence and reasoning. 

Our findings have several limitations. Our analysis covers a single platform, and MAHA-aligned discourse on the X, Telegram, and Meta platforms may exhibit different patterns. Additionally, Reddit's content moderation likely removed the most extreme MAHA-aligned posts before our data collection, which may cause our pipeline to understate the prevalence and intensity of pro-MAHA stances. Platforms with weaker moderation could, therefore, yield different stance distributions, making cross-platform comparison an important direction for future work.
Our keyword-based filtering, despite being two-stage and validated, may miss MAHA-aligned engagement expressed implicitly --- for example, short reactive comments (e.g., "agree" or "disagree") that lack explicit MAHA vocabulary. More sophisticated context-aware classification in the early stages could recover these cases. 
The within-MAHA clusters we report are exploratory and depend on the specific 12 themes we selected; alternative theme definitions could yield different sub-structures. 
Finally, the LIWC effect sizes are small to moderate, which is typical for matched-corpus comparisons at this scale; however, we showed the convergence of LIWC patterns with example-post evidence.

These findings matter for public-health communication and content moderation. MAHA functions as a connected belief system that grows through identifiable entry points and is expressed through a distinct rhetorical style. Interventions that target individual topics in isolation are unlikely to be effective; addressing the movement requires engaging its overall structure.
\subsection{Acknowledgment}
A generative AI tool (Claude) was used to support code and figure development, and editing of prose. The authors designed the study, performed all analyses, interpreted results, and prepared the final text.
\bibliography{bibtex}
\subsubsection{Ethical Statement}
All data analyzed in this study comes from Reddit's publicly accessible submissions and comments, accessed in accordance with the platform's terms. We report analyses exclusively at the aggregate community level; no usernames or user identifiers are released. 
This study was not subject to IRB review as it involves analysis of publicly available data with no direct interaction with human subjects.

\end{document}